\documentclass[a4paper,11pt]{article}
\usepackage[dvipdfmx]{graphicx}

\usepackage{jcappub} 
\usepackage[T1]{fontenc} 
\usepackage{comment}
\usepackage{subfigure}
\title{Experimental Search for Hidden Photon CDM in the eV mass range with a Dish Antenna}

\author[a]{J. Suzuki,}
\author[a]{T. Horie,}
\author[b]{Y. Inoue,}
\author[a,c]{and M. Minowa}

\affiliation[a]{Department of Physics, School of Science, The University of Tokyo,\\7-3-1 Hongo, Bunkyo-ku, Tokyo 113-0033, Japan}
\affiliation[b]{International Center for Elementary Particle Physics, The University of Tokyo,\\7-3-1 Hongo, Bunkyo-ku, Tokyo 113-0033, Japan}
\affiliation[c]{Research Center for the Early Universe (RESCEU), School of Science, The University of Tokyo,\\7-3-1 Hongo, Bunkyo-ku, Tokyo 113-0033, Japan}

\emailAdd{jsuzuki@icepp.s.u-tokyo.ac.jp}

\abstract{
A search for hidden photon cold dark matter (HP CDM) using 
a new technique with a dish antenna is reported. 
From the result of the measurement, we found no evidence for the existence of HP CDM 
and set an upper limit on the photon-HP mixing parameter $\chi$ of $\sim 6\times 10^{-12}$ 
for the hidden photon mass $m_\gamma = 3.1 \pm 1.2\,\rm{eV}$.
}

\begin{document}
\maketitle
\flushbottom

\section{Introduction}
\label{sec:intro}

A lot of evidence from observations of astronomical sources indicates the existence of invisible non-baryonic matter (dark matter, DM) in the universe. 
Probing the nature of dark matter is one of the central issues in astrophysics and cosmology today, 
and a variety of efforts have been paid to directly detect dark matter particles. 

The most prominent candidate for dark matter is the weakly interacting massive particle (WIMP), 
and most of the current experiments aim to detect 
WIMPs via their elastic scattering off atomic nuclei. 
However, there are alternative candidates to account for the features of DM, 
and Weakly Interacting Slim Particles (WISP), e.g. axion-like particles (ALP)  or hidden-sector photons (HP), can be the main component of DM~\cite{Arias}.

Hidden photons ($\tilde{X}^\mu$) are 
light extra U(1) gauge bosons 
which have kinetic mixing with the ordinary photons~\cite{Holdom}. 
The low energy Lagrangian of this model reads:
\[
\mathcal{L}=-\frac{1}{4}F_{\mu\nu}F^{\mu\nu}-\frac{1}{4}\tilde{X}_{\mu\nu}\tilde{X}^{\mu\nu}-\frac{\chi}{2}F_{\mu\nu}\tilde{X}^{\mu\nu}+\frac{m_{\gamma'}^{2}}{2}\tilde{X}_{\mu}\tilde{X}^{\mu}+J^{\mu}A_{\mu}, 
\]
where $F_{\mu\nu}$ is the field strength of the ordinary electromagnetic field $A^\mu$, $\tilde{X}_{\mu\nu}$  the field strength of the HP field $\tilde{X}^\mu$, $m_{\gamma '}$ the mass of the hidden photon, and $\chi$ the mixing parameter. 
Recently, it has been shown that a huge region in the parameter spaces spanned by $\chi$ and $m_{\gamma}$ can explain the observed CDM~\cite{Arias}.

This hidden photon CDM scenario 
can be experimentally investigated 
via photon-HP kinetic mixing $(\chi/2)F_{\mu\nu}\tilde{X}^{\mu\nu}$. 
For example, the Axion Dark Matter eXperiment (ADMX)~\cite{ADMX_detector}, 
which employs a resonant cavity and magnetic field to search for axion dark matter, 
also has sensitivity to hidden photon CDM, 
and upper limit for the kinetic mixing parameter $\chi$ was calculated in Ref.~\cite{Arias} 
from the previous results for axion DM~\cite{ADMX_result1, ADMX_result2, ADMX_result3, ADMX_result4, ADMX_result5}.

Additionally, another novel method with a spherical mirror
was recently proposed~\cite{Horns}, with which 
wider mass-range can be probed without rearranging the setup. 
In this method, 
ordinary photons of energy $\omega \simeq m_{\gamma '}$ induced by HP DM via kinetic mixing 
are emitted in the direction perpendicular to the surface of the mirror, 
resulting in concentration of the power to the center of the mirror sphere.
Assuming that DM is totally made up of hidden photons, 
the power concentrated on the center of the spherical mirror is
\[
	P=2\alpha^{2}\chi^{2}\rho_{{\rm CDM}}A_{{\rm dish}},
\]
where $\alpha=\cos\theta$ with $\theta$ the angle between the direction of HP vector and the surface of the mirror, 
$\rho_{\rm{CDM}} = (m_{\gamma '}^2/2)|X|^2\simeq 0.3\,\rm{GeV/cm^{3}}$ the energy density of CDM,
and $A_{\rm{dish}}$ the area of the mirror.
Solving this equation for $\chi$, we obtain the sensitivity to $\chi$~\cite{Horns}, 
\begin{equation}
\chi_{{\rm sens}}=5.6\times10^{-12}\left(\frac{R_{\gamma,\,\rm{det}}}{1\,{\rm Hz}}\right)^{\frac{1}{2}}\left(\frac{m_{\gamma'}}{{\rm eV}}\right)^{\frac{1}{2}}\left(\frac{0.3\,{\rm GeV/cm^3}}{\rho_{{\rm CDM,\, halo}}}\right)^{\frac{1}{2}}\left(\frac{1\,{\rm m^{2}}}{A_{{\rm dish}}}\right)^{\frac{1}{2}}\left(\frac{\sqrt{2/3}}{\alpha}\right),
\label{eq:limit}
\end{equation}
where $R_{\gamma,\,\rm{det}}$ is the minimum count rate which can be detected by the detector.
$R_{\gamma,\,\rm{det}}$ depends on the duration of the measurement $T$, the dark count rate of the detector $\nu$, and the quantum efficiency of the detector $\eta$ for the optical set-up.
Expressing $R_{\gamma,\,\rm{det}}$ in those values
yields
\[
\begin{split}
\chi<&\,5.5\times10^{-13}\times\left(\frac{0.1}{\eta}\right)^{\frac{1}{2}}\left(\frac{\nu}{{\rm Hz}}\frac{100\,{\rm day}}{T}\right)^{\frac{1}{4}}\left(\frac{m_{\gamma'}}{{\rm eV}}\right)^{\frac{1}{2}}\\
&\,\times	\left(\frac{0.3\,{\rm GeV/cm^{3}}}{\rho_{{\rm CDM,\, halo}}}\right)^{\frac{1}{2}}\left(\frac{1\,{\rm m^{2}}}{A_{{\rm dish}}}\right)^{\frac{1}{2}}\left(\frac{\sqrt{2/3}}{\alpha}\right)\,(95\%\,{\rm CL}),
\end{split}
\]
which suggests that high quantum efficiency and large area of the mirror is required for high sensitivity.

In spite of the importance to explore HPDM scenario and 
relatively easy design of the `dish' method, 
there have been no published experiments employing it.
In this paper, 
we report the first search for hidden photon CDM using a dish antenna.

\section{Experimental apparatus}
\begin{figure}
\begin{center}
	\includegraphics[width=0.8\textwidth]{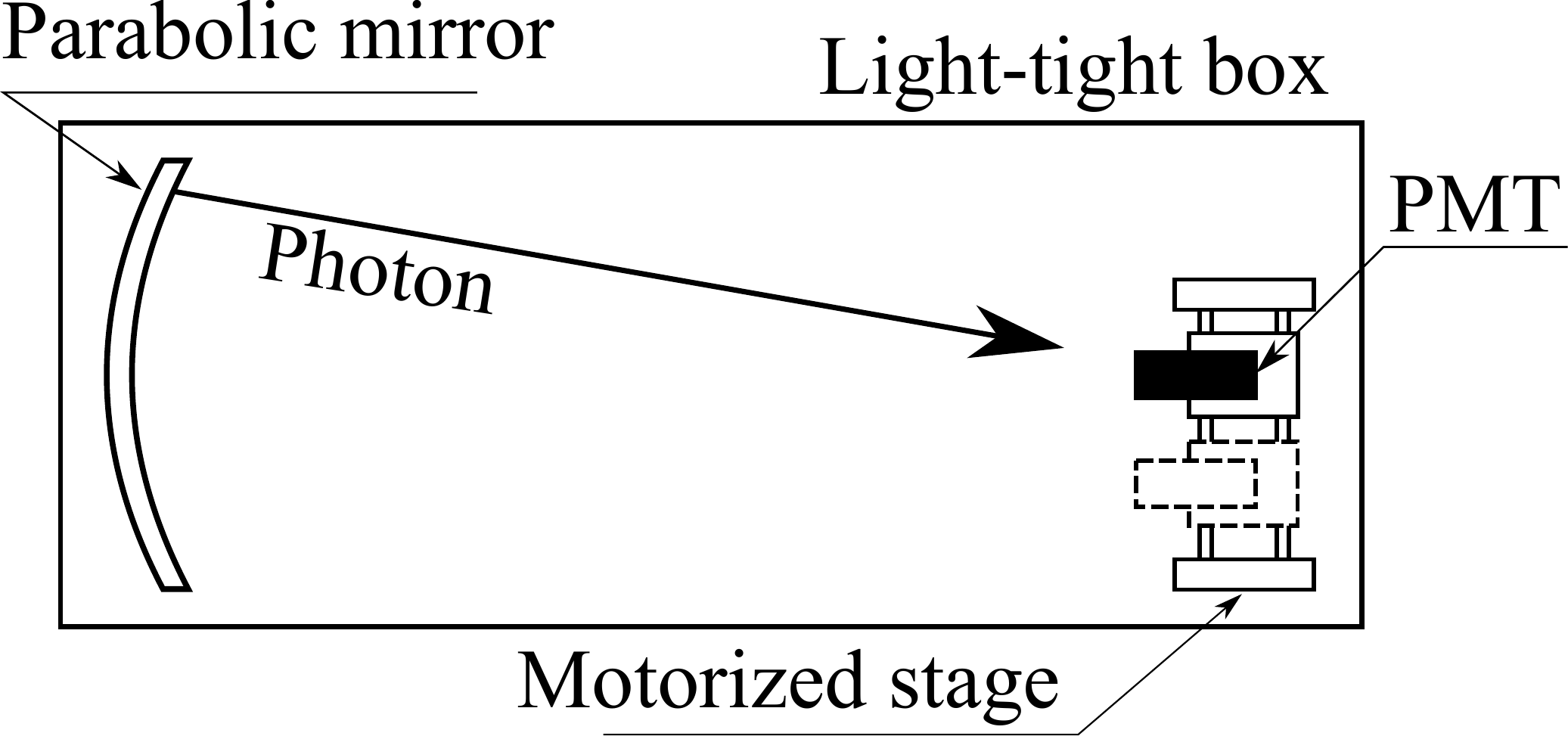}
\end{center}
	\caption{\label{fig:sch} Schematic diagram of the apparatus. 
Photons from non-relativistic HPs are emitted from the surface of the mirror and converge to a small region at twice the focal length of the mirror. 
A photon-counting PMT was placed at the point of convergence and detects emitted photons. 
The PMT is mounted on a motorized stage, which shifts the position of the detector to measure background noise. 
Devices described above are installed in a light-tight box to shield from the ambient light.
}
\end{figure}

A schematic diagram of the apparatus is given in  Fig.~\ref{fig:sch}. 
Non-relativistic HPs near the surface of a reflector 
induce emission of photons in the direction perpendicular to the surface. 
A photodetector is placed at the point of convergence and detect emitted photons. 
The detector is mounted on a motorized stage. 
Devices described above are enclosed in a light-tight box of 1m $\times$ 1m $\times$ 3m to prevent ambient light from entering to the photodetector. 

We used a parabolic mirror as a `dish' in the method. 
The parabolic mirror was previously used in the solar HP search experiment~\cite{Mizumoto}, 
and is 500mm in diameter, 19mm thick, 1007mm focal length and the focal spot diameter is 1.5mm. 
The reflectance of the mirror was measured by the manufacturer as a function of the wavelength, 
and was higher than 80\% over the range of interest. 
We used a parabolic surface instead of a spherical surface originally proposed in Ref. ~\cite{Horns}.
From the diameter and the focal length of the parabolic mirror, photons emitted perpendicularly to the surface 
are calculated to concentrate to a small area of 4 mm in diameter at twice the focal length of the mirror, 
which is small enough compared to the effective area of the photodetector described below. 

A photomultiplier tube (PMT) was employed as the detector of emitted photons. 
We selected Hamamatsu Photonics R3550P because of its low dark count rate of $\sim 5 \rm{Hz}$. 
This PMT has a low noise bialkali photocathode whose effective area is 22 mm in diameter and has sensitivity for photons of wavelength range 300--650 nm with a peak quantum effciency of 17 \%.
As described below, 
Cherenkov light from cosmic-ray muons 
can be a major source of systematic error. 
To diminish the effect, 
we limited the effective area of the PMT to 11 mm in diameter by a black paper shield. 

The optical system was aligned 
with an accuracy sufficient for 
emitted photons to converge 
within the limited region of the effective area of the PMT, 
taking into account the deviation and the dispersion 
originated from the velocity distribution
\footnote{We adopted the isothermal model for the distribution of velocities, 
which is widely accepted for a working hypothesis 
in the field of the direct detection of dark matter. }
of the dark matter particles.

We used a motorized stage to shift the position of the PMT, which enabled us to measure background noise. 
The stage is driven by a stepper motor connected to a ball screw mechanism. 
We used a pair of mechanical micro-switches to calibrate the position of the stage even inside the light-tight box. 
The stepper motor was controlled from a computer, which recorded the time when the switch was turned on and when the stage started to move.

The mirror and the detector were mounted on a steel frame, which rigidly holds the arrangement. 
We confirmed the stability of the optical alignment by checking the position of the detector before and after a month of test runs. 
The frame 
is wrapped with 150$\mu\rm{m}$ thick black polyethylene sheets.
The whole architecture was placed in the underground laboratory of the University of Tokyo 
at East longitude 139$^\circ$45'47'', North latitude 35$^\circ$42'50'', 
and the mirror is directed to the West.

\begin{figure}
\begin{center}
	\includegraphics[width=0.8\textwidth]{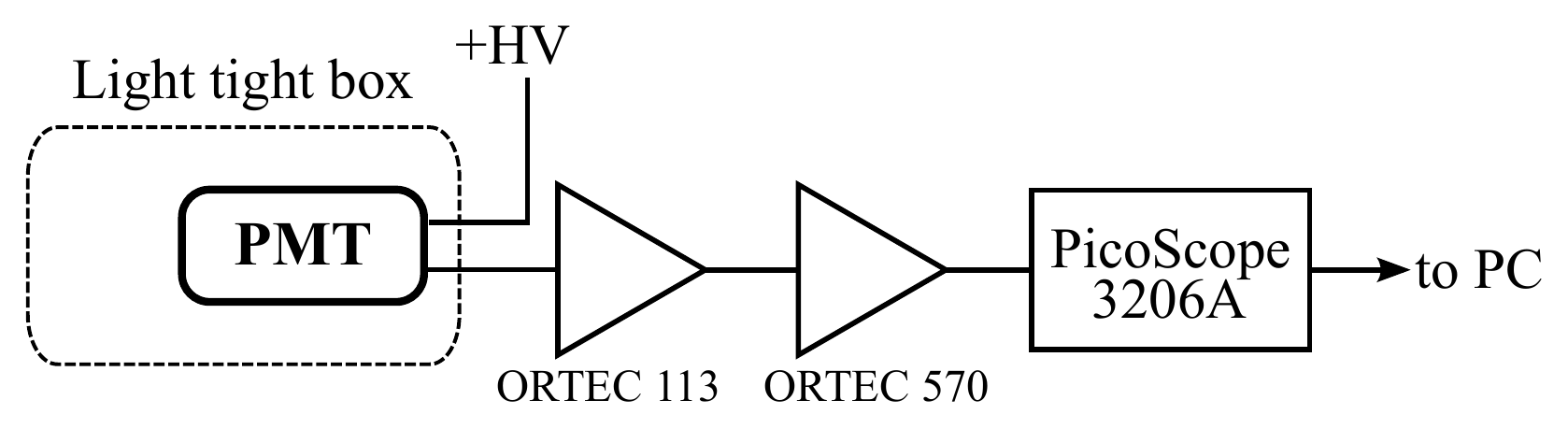}
\end{center}
	\caption{\label{fig:daq} Schematic diagram of the data aquisition system.
The PMT output current is connected to a charge-sensitive preamplifier (ORTEC 113) 
followed by a shaper (ORTEC 572), then digitized by an oscilloscope (PicoScope 3206A).
}
\end{figure}

Figure~\ref{fig:daq} shows a schematic diagram of the data acquisition system. 
The PMT output current is connected to a charge-sensitive preamplifier (ORTEC 113) followed by a shaper (ORTEC 572). 
The signal is then sent to a digital oscilloscope (PicoScope 3206A), 
which samples the signal at 10M samples/sec and streams the data to a computer through USB 2.0. 
Event triggering and pulse-height analysis are done in a software, which records pulse heights and  arrival times. 

\section{Measurement and analysis}

The existence of HP CDM would yield a single photon and be detected by the PMT as a single-photon event. 
We calibrated the PMT with a pulsed very faint blue LED light 
to study its response to a single photo-electron. 
The pulse-height spectrum constructed from the calibration data 
was fitted by a model function which is the sum of a Gaussian and a exponential curve, 
and the result was later used as the template 
in the analysis of HP CDM search. 

The data were recorded from February 2015 to March 2015. 
The overall duration of the measurement was $8.3\times10^5 \rm{s}$ in total with the PMT 
at the position of convergence of the HPDM signal (signal, S) 
and at the position displaced by 25 mm from position S (background, B). 
The motorized stage displaced the PMT between position S and position B. 
The measurement was performed as follows: 
(i) acquire the data at position S for 30 seconds, 
(ii) shift the position of the PMT from S to B, 
(iii) acquire the data at position B for 30 seconds, 
(iv) restore the position from B to S. 
We repeated the procedure to reduce the effect of the change in temperature of the PMT as described bellow.

\begin{figure}
	\centering
	\includegraphics[width=0.8\textwidth]{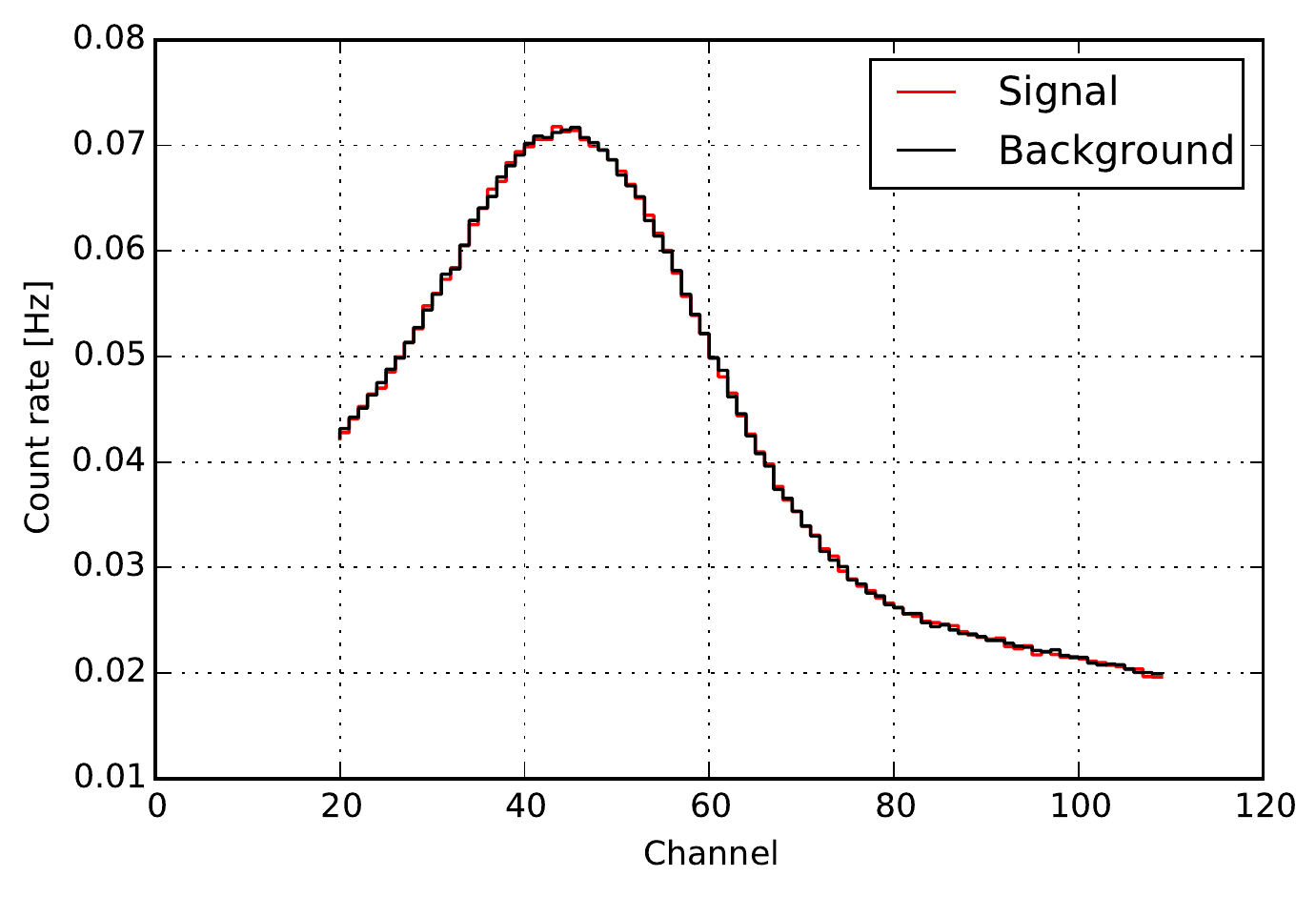}\\
	\includegraphics[width=0.8\textwidth]{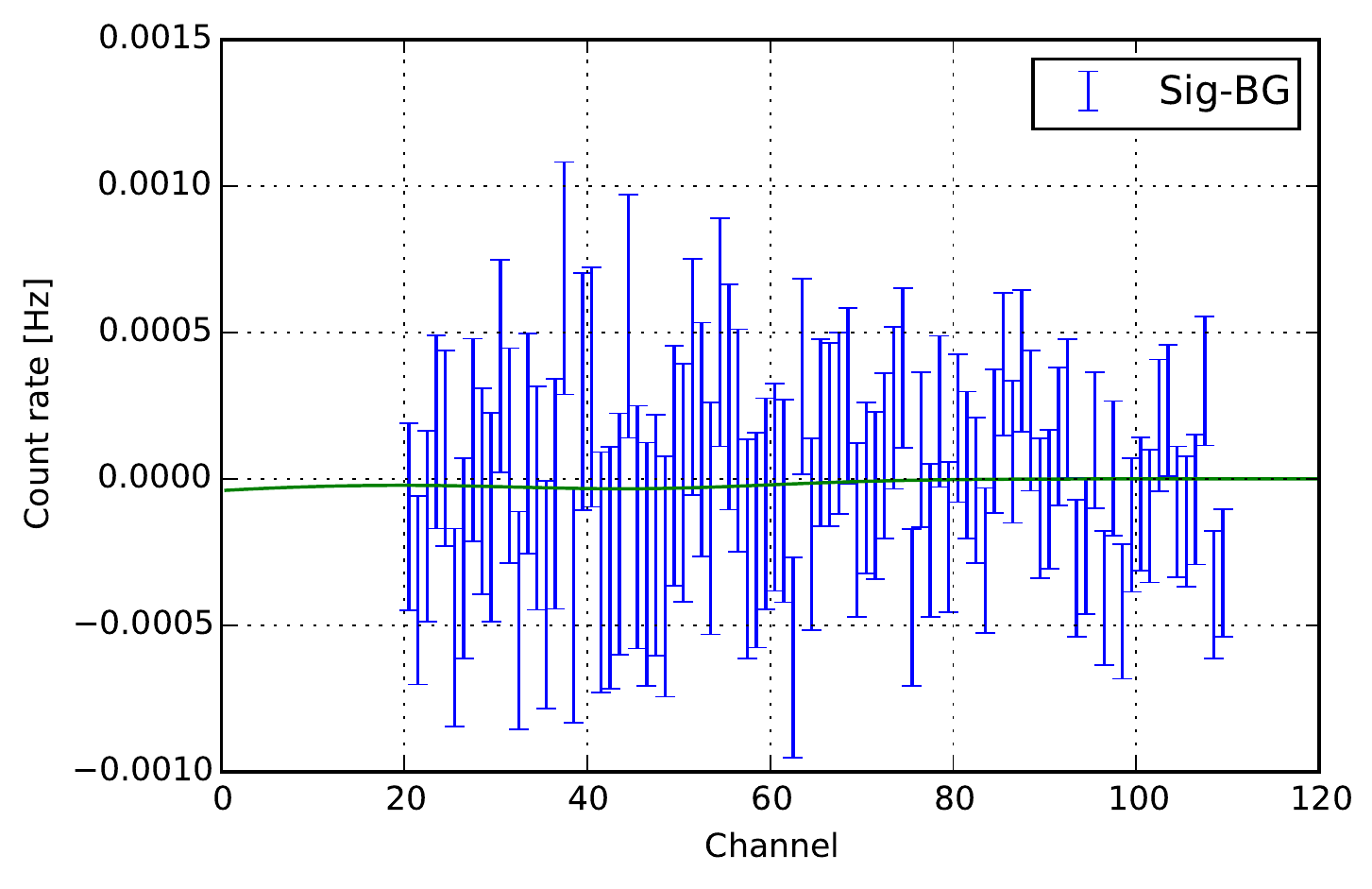}
	\caption{
({\bf top}) Pulse height spectra taken at S (red) and at B (black) 
normalized by the live times. 
({\bf bottom}) Difference between spectra taken at S and at B (blue), 
fitted with the template function 
obtained in the calibration utilizing LED light (green).
}
	\label{fig:Hists} 
\end{figure}

The final result is shown in Fig.~\ref{fig:Hists}.
The upper figure shows 
pulse height spectra with the PMT both at positions S and B divided by the duration of the measurements. 
Those two spectra almost coincide with each other, 
from which we can conclude that their main constituents are dark counts.
In the bottom figure, the spectrum at B was subtracted from the spectrum at S 
to extract the possible signal of the existence of HPDM, 
then the result of subtraction was fitted with the template function obtained in the analysis for LED  pulses. 
The difference of the count rate 
between at S and at B 
is then estimated to be 
\begin{equation}
\label{eq:fit_result}
N_{\rm{fit}}=(-1.9\pm3.8)\times10^{-3}\,\rm{Hz}
\end{equation}
after proper normalization.

The main sources of systematic errors are 
(i) temperature dependence of the dark count rate of the PMT and 
(ii) Cherenkov emission in the window of the detector.

\begin{figure}
\begin{center}
	\includegraphics[width=0.8\textwidth]{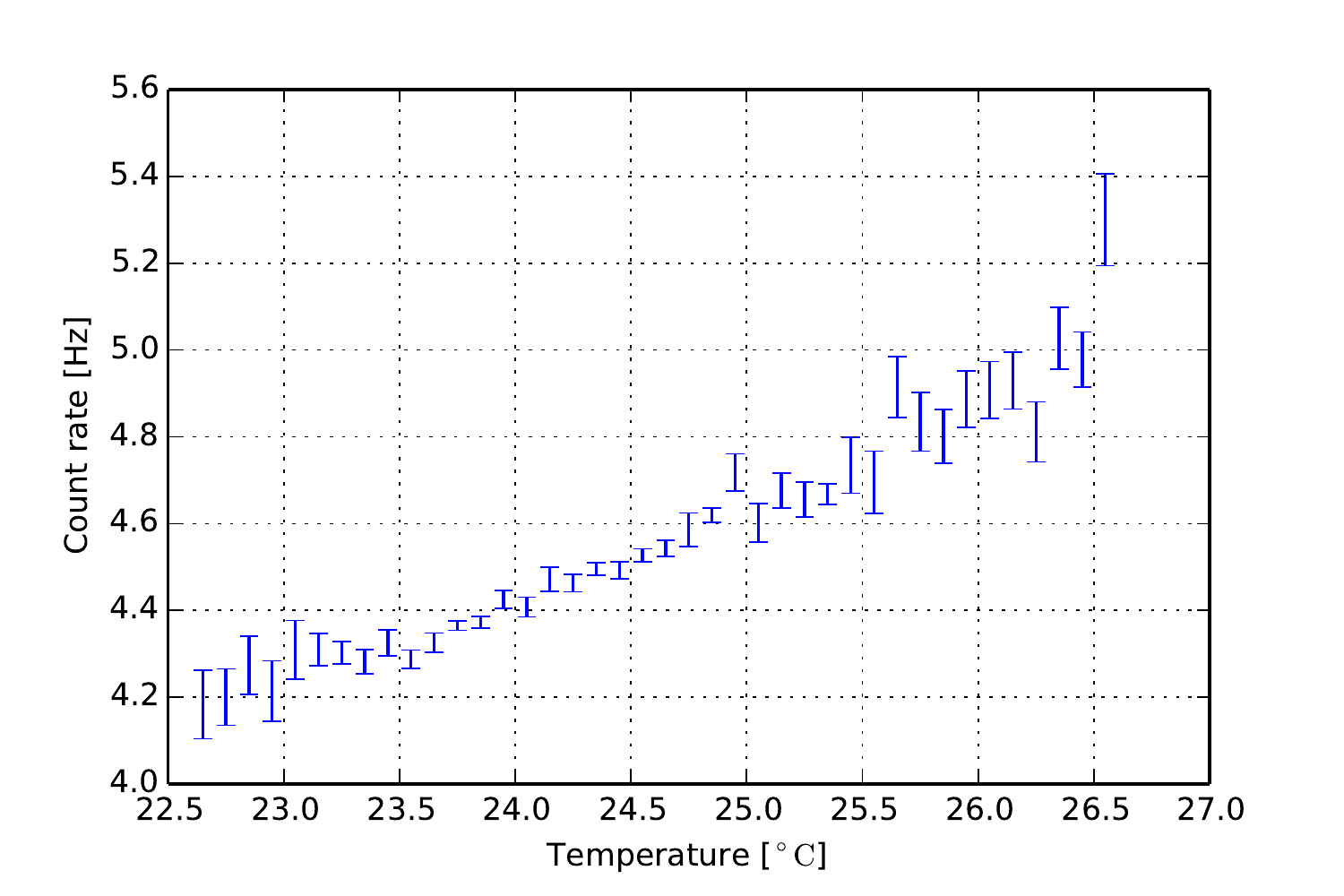}
\end{center}
	\caption{\label{fig:temp_dep} Temperature dependence of the dark count rate. 
The count rate for each temperature bin is plotted on the vertical axis, and the temperature is on the horizontal axis.
}
\end{figure}

The temperature dependence of the dark count rate 
was studied with a Pt100 thermometer located beside the detector. 
Figure~\ref{fig:temp_dep} shows the dependence of the dark count rate on temperature. 
We can see that the dark count rate rises linearly with respect to the temperature with coefficient of $0.21 \,\rm{Hz}/^{\circ}\rm{C}$. 
The measurement procedure described above, 
in which the signal and the background are acquired quasi-simultaneously, 
was intended to reduce 
the effect of the temperature variation between measurements. 
From this result 
and temperature monitoring during the whole measurement period, 
we estimated the systematic uncertainty due to the dependence of the dark count rate on temperature 
as $\pm 0.5\times10^{-3}\,\rm{Hz}$ in count rate. 

In addition to the effect of the temperature variation, 
the Cherenkov emission caused by cosmic-ray muons 
could be a source of systematic errors. 
Cherenkov photons would be emitted 
by cosmic-ray muons 
passing through the window of the PMT. 
If the PMT is located at position S, 
some portion of emitted photons would be reflected by the mirror 
and re-enter into the PMT 
because of the optical configuration. 
On the other hand, no Cherenkov light is reflected back to the PMT if it is at position B. 
The most pessimistic evaluation of this effect without the black paper shield 
yields a count rate of $\sim 0.01\,\rm{Hz}$, 
which would be larger than the statistical uncertainty 
for measurement over a week. 
In order to reduce this effect 
the effective area of the PMT 
was limited to 11 mm in diameter 
from its original size of 22mm in diameter. 
As the signal spot is well localized and 
the movement of the spot during a day caused by 
the dark-matter `wind'~\cite{Directional} 
is $\lesssim 2\,\rm{mm}$, 
the efficiency of the measurement therefore would not be affected by this treatment. 
The effect of the Cherenkov light to the final result 
is estimated to %
be less than 
$2\times10^{-3}\,\rm{Hz}$ in count rate with the limited effective area of the PMT, 
which potentially may cause a positive shift in the possible HPDM signal. 
However, we did not subtract this effect from $N_{\rm{fit}}$ in eq. (\ref{eq:fit_result}) and tried to estimate a conservative upper limit to it.

Combining eq. (\ref{eq:fit_result}) and the systematic error from the temperature variation, 
we obtain 
the possible count rate of the signal which originates from the existence of HPDM as 
\[
N=(-1.9\pm 3.8 \rm{(stat.)} \pm 0.5{(sys.)})\times10^{-3}\,\rm{Hz},
\]
which shows no significant evidence for the existence of HPDM. 
From this we calculate an upper limit of 
\[
N_{\rm UL95} = 6.4 \times 10^{-3}\,\rm{Hz}
\]
at 95\% confidence level.
Making use of this value of photons detected at the PMT 
divided by the quantum efficiency of the PMT and the reflectivity of the mirror, 
and according to eq.~(\ref{eq:limit}), 
the upper limit to the mixing parameter $\chi$ is calculated 
assuming that DM is dominated by HP and
that HPDM is not polarized and the direction is isotropically distributed in the celestial sphere, i.e. $\alpha=\sqrt{2/3}$.

\begin{figure}
	\centering
	\includegraphics[width=1.0\textwidth]{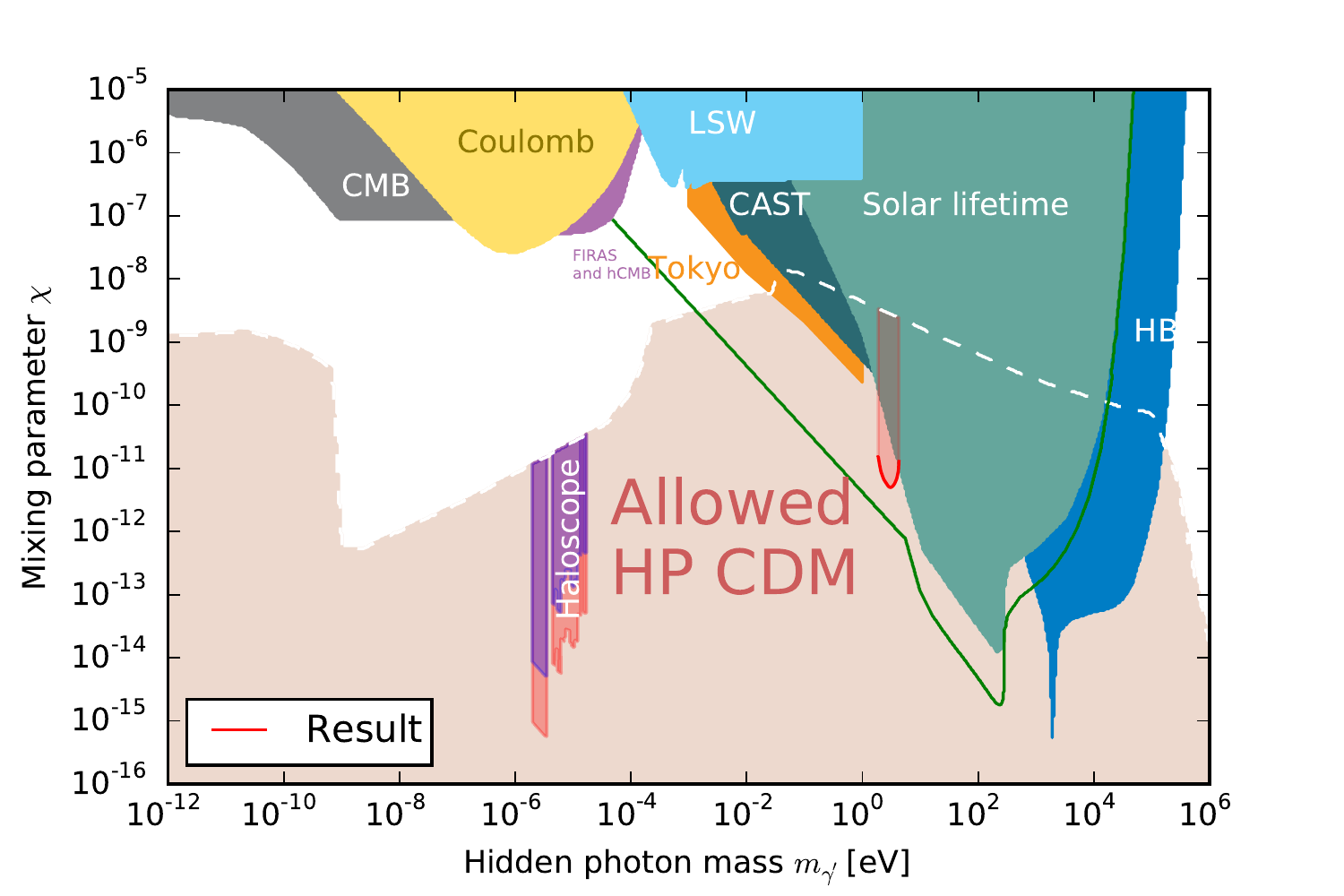}
	\caption{Excluded region of $\chi\hbox{--}m_{\gamma'}$ parameter space. 
The upper limit obtained in this experiment is shown in a solid red line.
The region allowed for HPDM ~\cite{Arias} is filled with light reddish brown, and the excluded region from this experiment is overpainted with transparent red. 
The limit translated from the previous results for axion DM with the assumption that DM is mainly composed of HP~\cite{Arias} is marked as ``Haloscope''. 
Other filled areas are excluded by other experiments or 
considerations on astronomical sources.
The regions excluded by 
precision tests of Coulomb's law~\cite{Williams, Bartlett}, 
``Light Shining through Walls'' experiments~\cite{ALPS, BMV, GammeV, LIPSS}, 
the CAST experiment~\cite{CAST}, 
solar hidden photon search utilizing the Tokyo Axion Helioscope~\cite{Mizumoto} 
and FIRAS CMB spectrum~\cite{FIRAS} 
are marked as ``Coulomb'', ``LSW'', ``CAST'', ``Tokyo'' and ``FIRAS'', respectively.
Constraints from the solar lifetime which takes only transverse mode into account~\cite{CAST} 
is marked as ``Solar lifetime'', 
and the calculations which properly deal with longitudinal mode of the massive state~\cite{An, RedRev} is shown with a green solid curve.
}
	\label{fig:limit_to_coupling}
\end{figure}

The result is shown in Fig.~\ref{fig:limit_to_coupling} with a red solid curve. 
The region allowed for HPDM ~\cite{Arias} is filled with light reddish brown, and the excluded region from this experiment is overpainted with transparent red. 
The limit translated from the previous results for axion DM with the assumption that DM is mainly composed of HP~\cite{Arias} is marked as ``Haloscope''. 
Other filled areas are excluded by other experiments or 
considerations on astronomical sources.
The regions excluded by 
precision tests of Coulomb's law~\cite{Williams, Bartlett}, 
``Light Shining through Walls'' experiments~\cite{ALPS, BMV, GammeV, LIPSS}, 
the CAST experiment~\cite{CAST}, 
solar hidden photon search utilizing the Tokyo Axion Helioscope~\cite{Mizumoto} 
and FIRAS CMB spectrum~\cite{FIRAS} 
are marked as ``Coulomb'', ``LSW'', ``CAST'', ``Tokyo'' and ``FIRAS'', respectively.
Constraints from the solar lifetime which takes only transverse mode into account~\cite{CAST} 
is marked as ``Solar lifetime'', 
and the calculations which properly deal with longitudinal mode of the massive state~\cite{An, RedRev} is shown with a green solid curve.

Although the upper limit for $\chi$ obtained in this experiment 
is nominally weaker than the constraint from the solar lifetime, 
it is still significant because the calculation of the limit from the solar lifetime strongly depends on the solar model, 
in which severe discrepancy with the real situation can occur,
while our experimental limit only assumes that DM is mainly composed of hidden-sector photons.
This work also showed an example of detailed experimental method with points to make note of to search for HPDM with a dish antenna.

\section{Conclusion}

The experimental search for HP CDM in the eV-mass region was performed for the first time using the novel technique with a dish antenna. 
No excess of count rate was observed in the exposure for  $8\times 10^{5}$ s, and  the limit for the mixing parameter $\chi$ was calculated assuming that dark matter is dominated by HPs. 

\section*{Acknowledgments}

T. Horie acknowledges support by Advanced Leading Graduate Course for Photon Science (ALPS) at the University of Tokyo.  
This reaserch is supported by the Grant-in-Aid for challenging Exploratory Research by MEXT, Japan, 
and also by the Research Center for the Early Universe, School of Science, the University of Tokyo. 


\end{document}